\documentclass[aps,prb,floatfix,twocolumn,reprint,amsmath,amssymb,superscriptaddress,longbibliography]{revtex4-2}
\usepackage{amsmath,amssymb,amsfonts,mathrsfs,braket,amsthm,bm,graphicx,booktabs,xcolor,siunitx}
\usepackage{amsthm}
\usepackage{CJK}
\usepackage{bm}
\usepackage{mathdots}
\usepackage{lipsum}
\usepackage{babel}
\usepackage{graphicx}
\usepackage{makecell}
\usepackage{booktabs}
\usepackage{multirow}
\usepackage{gensymb}
\usepackage{rotating}
\usepackage{float}
\usepackage{braket}
\usepackage[breaklinks]{hyperref}
\usepackage{color}
\usepackage{array}
\hypersetup{colorlinks=true, linkcolor=blue, citecolor=blue, filecolor=blue, urlcolor=blue}
\usepackage[version=4]{mhchem}

\newcommand{\mi}{\mathrm{i}}
\newcommand{\me}{\mathrm{e}}

\allowdisplaybreaks

\begin{document}

\title{Chiral Superconductivity in Periodically Driven Altermagnet/Superconductor Heterostructures}

\author{Xiaolin Wan}
\affiliation{Institute for Structure and Function $\&$ Department of Physics, Chongqing University, Chongqing 400044, P. R. China}
\affiliation{Chongqing Key Laboratory for Strongly Coupled Physics,  Chongqing 400044, P. R. China}

\author{Zheng Qin}
\affiliation{Institute for Structure and Function $\&$ Department of Physics, Chongqing University, Chongqing 400044, P. R. China}
\affiliation{Chongqing Key Laboratory for Strongly Coupled Physics,  Chongqing 400044, P. R. China}

\author{Fangyang Zhan}
\affiliation{Institute for Structure and Function $\&$ Department of Physics, Chongqing University, Chongqing 400044, P. R. China}
\affiliation{Chongqing Key Laboratory for Strongly Coupled Physics, Chongqing 400044, P. R. China}
\affiliation{Center for Quantum materials and devices, Chongqing University, Chongqing 400044, P. R. China}

\author{Junjie Zeng}
\email[]{jjzeng@cqu.edu.cn}
\affiliation{Institute for Structure and Function $\&$ Department of Physics, Chongqing University, Chongqing 400044, P. R. China}
\affiliation{Chongqing Key Laboratory for Strongly Coupled Physics, Chongqing 400044, P. R. China}
\affiliation{Center for Quantum materials and devices, Chongqing University, Chongqing 400044, P. R. China}

\author{Dong-Hui Xu}
\email[]{donghuixu@cqu.edu.cn}
\affiliation{Institute for Structure and Function $\&$ Department of Physics, Chongqing University, Chongqing 400044, P. R. China}
\affiliation{Chongqing Key Laboratory for Strongly Coupled Physics, Chongqing 400044, P. R. China}
\affiliation{Center for Quantum materials and devices, Chongqing University, Chongqing 400044, P. R. China}

\author{Rui Wang}
\email[]{rcwang@cqu.edu.cn}
\affiliation{Institute for Structure and Function $\&$ Department of Physics, Chongqing University, Chongqing 400044, P. R. China}
\affiliation{Chongqing Key Laboratory for Strongly Coupled Physics, Chongqing 400044, P. R. China}
\affiliation{Center for Quantum materials and devices, Chongqing University, Chongqing 400044, P. R. China}

\begin{abstract}
    The interplay between magnetism and superconductivity provides a fertile ground for engineering exotic topological phases, while dynamical control via periodic driving offers a unique avenue to access quantum states that are inaccessible in static equilibrium. Here, we propose a strategy to achieve the Floquet chiral topological superconductivity in an altermagnet-superconductor heterostructure driven by elliptically polarized light. We show that for $s$-wave pairing, the system undergoes a transition from a trivial to a chiral topological superconducting phase. More strikingly, with the introduction of mixed $s+d$-wave pairing, we find that the system can access Floquet chiral topological superconducting phases with highly tunable Chern numbers up to $|\mathcal{N|}=4$. These exotic phases are attributed to the intertwining of altermagnetism, superconducting pairing, and the periodic driving field. Our work establishes the light-driven altermagnetic heterostructure as a versatile platform for exploring and manipulating high-Chern-number chiral topological superconductivity.
\end{abstract}

\maketitle

\textit{Introduction}---The exploration of topological superconductors (TSCs) attracts intensive interest in the modern condensed-matter community, as it can give rise to topologically protected Majorana fermions (also referred to as Majorana modes)~\cite{Sato2017,RevModPhys.83.1057,doi:10.7566/JPSJ.85.072001,Alicea2012,RevModPhys.87.137,Beenakker2013,Mandal2023,Kallin2016}. In particular, a chiral TSC with time-reversal symmetry breaking is one of the most fascinating topics due to the coexistence of band topology and unconventional pairing symmetries~\cite{Kallin2016,Han2025,PhysRevB.92.064520,PhysRevLett.117.047001,PhysRevLett.132.036601,RevModPhys.80.1083,PhysRevB.61.10267,PhysRevB.82.184516}. The chiral TSCs exhibit a full pairing gap in the two-dimensional (2D) bulk and possess $\mathcal{N}$ gapless chiral Majorana modes at the edge (where $\mathcal{N}$ is the BdG Chern number). Of particular importance are the chiral TSCs with an odd $\mathcal{N}$. The vortex bound states of these chiral TSCs can support a single Majorana zero mode, serving as fundamental building blocks for topological quantum computation~\cite{PhysRevLett.86.268,RevModPhys.80.1083,PhysRevB.61.10267,PhysRevB.95.235305,Lian2018,PhysRevB.99.195137,PhysRevB.82.184516}. Therefore, extensive studies have been carried out to explore chiral TSC candidates~\cite{doi:10.1126/science.aag2792,doi:10.1126/sciadv.aax9480,Ran2019,Jiao2020,Ming2023,Schemm2014,Jang2011,doi:10.1126/sciadv.1602579,Can2021,PhysRevResearch.2.032023,PhysRevLett.114.236803,PhysRevLett.113.097001,PhysRevB.87.180503,PhysRevLett.104.040502}; however, naturally occurring chiral TSCs remain scarce. In the past decade, many proposals have been raised to achieve chiral topological superconductivity in artificial heterostructures through the proximity coupling conventional $s$-wave superconductors~\cite{PhysRevLett.100.096407,PhysRevLett.105.077001,PhysRevLett.104.040502,PhysRevB.82.184516,Kezilebieke2020}. In this context, an external magnetic field or a ferromagnetic (FM) substrate is generally regarded as an essential factor for breaking time-reversal symmetry. Nevertheless, FM orders tend to suppress the proximity-induced superconducting gap, thereby posing challenges in achieving robust chiral topological superconductivity in magnet-superconductor heterostructures~\cite{doi:10.1126/science.275.5307.1767,RevModPhys.77.935,PhysRev.135.A550,RevModPhys.77.1321,PhysRevLett.86.2427,PhysRevB.38.8823}.

Recently, altermagnetism, an emerging magnetic phase characterized by alternating nonrelativistic momentum-dependent spin splitting while maintaining a vanishing net magnetization, has become the subject of intense recent studies~\cite{PhysRevX.12.031042,PhysRevX.12.040501,Smejkal2020,Hayami2019,Naka2019,PhysRevB.102.014422,Ma2021,PhysRevMaterials.5.014409,Krempasky2024,PhysRevX.12.040002}. Considering that the interplay between magnetism and superconductivity facilitates the emergence of unconventional superconductivity, the exploration of TSCs in an  altermagnetic (AM) system is an interesting avenue. In this regard, several AM-based strategies for designing topological superconductivity with vanishing net magnetization have been proposed~\cite{PhysRevB.110.205120,PhysRevB.109.224502,PhysRevB.108.205410,PhysRevLett.131.076003,PhysRevB.108.075425,g17w-xs73,3k12-2467,PhysRevB.109.134511,PhysRevB.108.054511,4318-ttvf,PhysRevB.110.165141,PhysRevLett.133.106601,PhysRevB.108.184505}. Strikingly, it has been demonstrated that altermagnetism is beneficial for realizing chiral Majorana fermions~\cite{PhysRevLett.133.106601,PhysRevB.108.184505,4318-ttvf,PhysRevB.110.165141}.

 It is well-known that Floquet engineering, owing to its dynamic manipulation on ultrafast timescales, offers a powerful tool for designing novel quantum phases with intriguing topological features~\cite{PhysRev.138.B979,PhysRevA.7.2203,PhysRevX.4.031027,Eckardt2015,Bukov04032015,Oka2019}. Periodic driving can enable promising Floquet Majorana modes beyond the conventional Floquet band topology~\cite{4tng-rhc4,PhysRevB.99.014301,PhysRevB.95.155407,PhysRevB.87.115420,Wang2026AdvSci,PhysRevB.103.085413,PhysRevLett.111.136402,PhysRevB.95.134508,PhysRevB.90.205127,PhysRevB.107.035427,Claassen2019,PhysRevLett.106.220402,PhysRevB.87.201109,PhysRevResearch.2.013124}, and thus numerous exotic phenomena related to Floquet TSCs have spurred extensive research activity~\cite{PhysRevLett.106.220402,PhysRevB.87.201109,PhysRevResearch.2.013124,PhysRevB.96.125144,PhysRevResearch.3.023108,PhysRevLett.111.047002,PhysRevB.99.094303,PhysRevLett.126.086801}. Moreover, an application of periodically-driven circularly polarized light (CPL) can introduce an effective time-reversal symmetry (TRS) breaking and flexibly switch the chirality of systems by changing the handedness of CPL; thus Floquet engineering can provide a promising platform to study and control the chiral Majorana modes~\cite{Claassen2019,Kitamura2022,Haxell2023,PhysRevB.96.195303,PhysRevLett.123.016806,PhysRevB.97.165142}. Since the intertwining between magnetism, topology, and superconductivity creates numerous opportunities for engineering and controlling a wide range of exotic phenomena, the exploration of topological superconductivity in AM systems via Floquet engineering is highly desirable.

\begin{figure}
    \centering
    \includegraphics[width=0.47\textwidth]{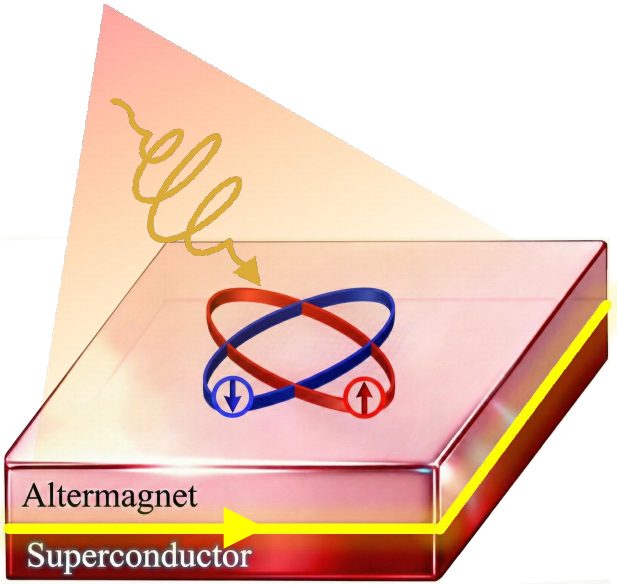}
    \caption{Schematic of the light-driven 2D AM-SC heterostructure, where the red and blue arrows denote spin-up and spin-down, respectively.
    \label{FIG1}}
\end{figure}

In this work, we theoretically investigate the topological phase transitions in an AM/SC heterostructure driven by elliptically polarized light (EPL), and propose a strategy for realizing the Floquet Chiral Topological Superconductor (FCTSC), as shown in Fig.~\ref{FIG1}. For the sake of concreteness, we focus on a heterostructure composed of a $ d $-wave AM and a SC with $ s $-wave or $ s + d $-wave pairing. We systematically calculate the Bogoliubov–de Gennes (BdG) Chern number $ \mathcal{N} $ to characterize the topological phases of the nonequilibrium steady states. We demonstrate that the light amplitude and the AM strength can synergistically promote the transition from a trivial SC into a topological phase hosting chiral Majorana fermions (CMFs). Remarkably, our calculations reveal that the driven system can host high Chern numbers ($ \mathcal{N} > 1 $), corresponding to multiple chiral Majorana edge modes. Our results establish a dynamical pathway to engineer and manipulate high-Chern-number topological phases in AM/SC platforms.

\textit{Lattice model and Floquet theory}.--- 
To investigate the topological phases, we employ a Bogoliubov–de Gennes (BdG) Hamiltonian constructed by adding superconducting pairing to the AM lattice model. The Hamiltonian for the AM/SC heterostructure explicitly includes several key components: the normal-state term, the Rashba spin-orbit coupling (SOC) term, the AM term, and two types of superconducting pairing terms — conventional $s$-wave pairing and unconventional $d_{x^2-y^2}$-wave pairing. 
The Hamiltonian of the system can be expressed as~\cite{PhysRevLett.133.106601}:
\begin{equation}\label{Eq.1}
    \begin{split}
        H_\text{BdG}(\bm{k}) &= H_0(\bm{k}) + \lambda_\text{R}(\bm{k}) + J_\text{AM}(\bm{k}) + \Delta(\bm{k})\tau^2\sigma^2 , \\
        H_0(\bm{k}) &= t (\cos k_x + \cos k_y - \mu) \tau^3, \\
        \lambda_\text{R}(\bm{k}) &= \lambda_\text{R} (\sin k_x \tau^3 \sigma^2 - \sin k_y \tau^0 \sigma^1), \\
        J_\text{AM}(\bm{k}) &= J_\text{AM} (\cos k_x - \cos k_y) \tau^3 \sigma^3, \\
        \Delta(\bm{k}) &= \Delta_0 + \Delta_1 (\cos k_x + \eta \cos k_y),
    \end{split} \tag{1}
\end{equation}
where $\tau$ and $\sigma$ denote Pauli matrices in Nambu and spin spaces, respectively. And $ t $ is the nearest-neighbor hopping strength, $ \lambda_\text{R} $ represents the Rashba SOC strength, and $ J_\text{AM} $ characterizes the AM strength. Furthermore, $ \Delta_0 $ and $ \Delta_1 $ are the superconducting pairing amplitudes for the $ s $-wave and $ d $-wave components, with $ \eta $ denoting the mixing parameter. 


\begin{figure}
    \centering
    \includegraphics[width=0.48\textwidth]{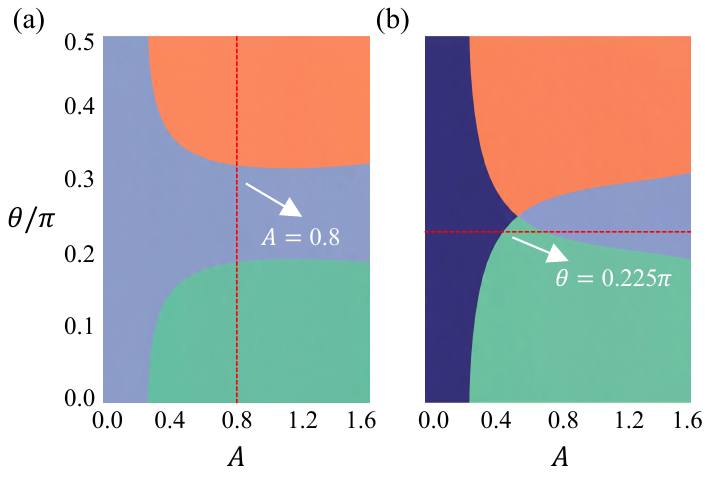}
    \caption{
    (a) Topological phase diagram characterized by Chern numbers in the $\theta$-$A$ parameter plane for an $s$-wave/AM heterojunction at $J_\text{AM} = 0.245$. The red dashed line indicates $A=0.8$. As $\theta$ varies along this dashed line, the system undergoes phase transitions between two distinct strong FCTSC phases and a topologically trivial phase.
    (b) Topological phase diagram in the $\theta$-$A$ parameter plane for an $s$-wave/AM heterojunction at $J_\text{AM} = 0.26$. As $A$ varies along the dashed line, the system evolves from a weak TSC state into an FCTSC state, and ultimately transitions into a topologically trivial state.
    The gray regions represent the topologically trivial phase. The yellow and green regions correspond to strong TSC states with Chern numbers $\mathcal{N} = \pm 1$, while the blue regions indicate a weak topological superconducting state with a Chern number of $\mathcal{N} = 0$.
    \label{FIG2}}
\end{figure}

A general form of the vector potential caused by the elliptically polarized light is given by
\begin{align}\label{Eq.2}
    \begin{split}
        \bm{A}(\tau) = \left(A_x \cos(\omega\tau), A_y \sin(\omega\tau), 0\right).
    \end{split}\tag{2}
\end{align}
The elliptically polarized driving field can be expressed as $ \bm{E}(\tau) = -\partial_\tau \bm{A}(\tau) $, where $ \bm{A}(\tau) = \bm{A}(\tau + T) $ is a time periodic, spatially homogeneous vector potential of period $ T = 2\pi/\omega $, and $\omega$ is the frequency of light. We incorporate the effect of the periodic driving field into the tight-binding Hamiltonian via the Peierls substitution, which introduces a time-dependent phase in the hopping amplitudes and is equivalent to the replacement $ \bm{k} \rightarrow \bm{k}+ \bm{A}(t)$. Setting $ \hbar = e = c = 1 $, this yields a Floquet system governed by a time-periodic Hamiltonian $ H(\tau) = H(\tau + T) $, which serves as the starting point for our high-frequency analysis. In the following discussion, we define the light field amplitude as $A=e\sqrt{A_{x}^{2}+A_{y}^{2}}$, with the polarization angle given by ${\theta}=\arctan(A_{y}/A_{x})$.

\begin{figure*}
    \centering
    \includegraphics[width=1 \textwidth]{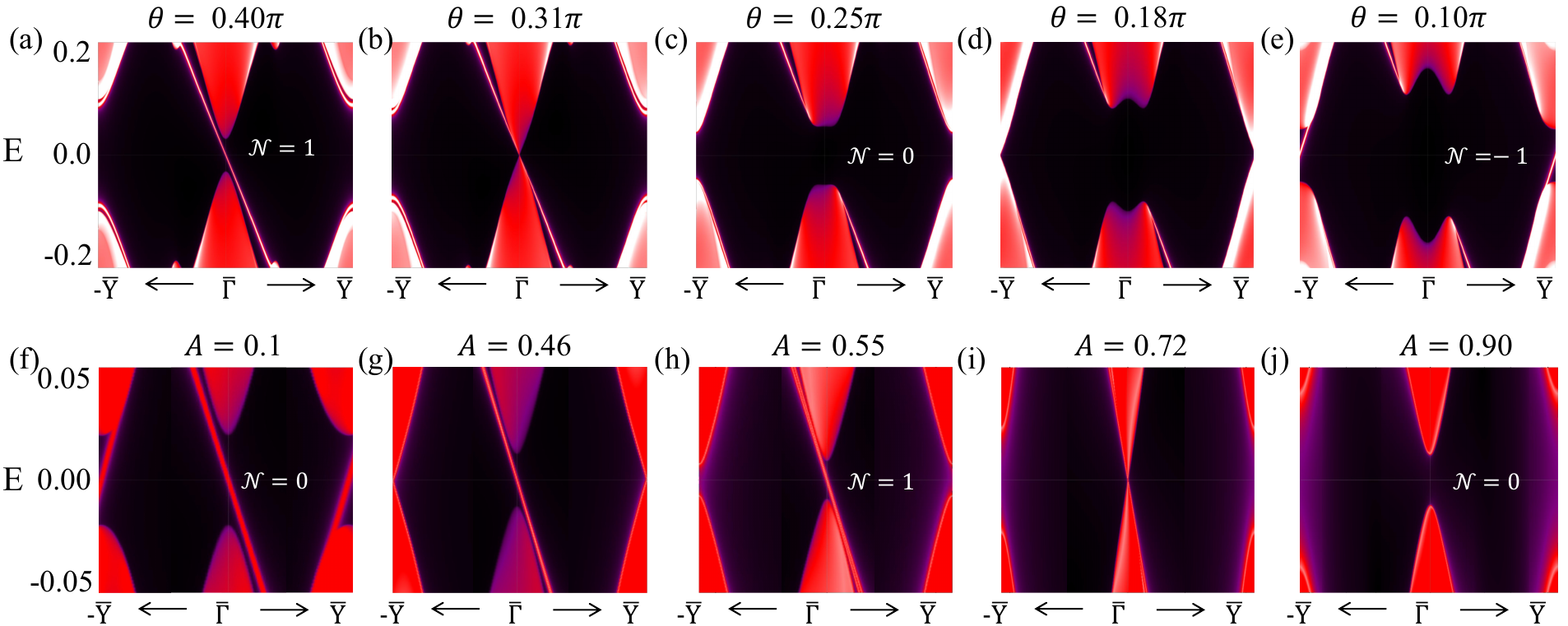}
    \caption{Edge states and topological phase transitions in an AM/s-wave SC system.
    (a)--(e) Edge state spectra along the $-\bar{\text{Y}} -\bar{\Gamma}-\bar{\text{Y}} $ path for fixed AM strength $J_\text{AM}=0.245$ and light amplitude $A=0.8$. The polarization angles correspond to $\theta = 0.1\pi$, $0.31\pi$, $0.25\pi$, $0.18\pi$, and $0.10\pi$, respectively. The corresponding Chern numbers are 
    $\mathcal{N}=1$, gap closing, $\mathcal{N}=0$, gap closing, and $\mathcal{N}=-1$.
    (f)--(g) Edge state spectra for fixed AM strength $J_\text{AM}=0.26$ and polarization angle $\theta=0.225\pi$. The light amplitudes are $A=0.3$, $0.46$, $0.55$, $0.72$, and $0.9$, respectively. The corresponding Chern numbers are $\mathcal{N}=0$, gap closing, $\mathcal{N}=1$, gap closing, and $\mathcal{N}=0$.
    Notably, in (f), two opposite chirality Majorana edge modes are present, resulting in a net Chern number of $\mathcal{N}=0$
    \label{FIG3}}
\end{figure*}

According to the Floquet theory~\cite{PhysRev.138.B979,PhysRevA.7.2203}, the time-dependent Schrödinger equation admits solutions of the form $ \ket{\Psi(\tau)} = \me^{-\mi\epsilon\tau}\ket{\Phi(\tau)} $, where $ \epsilon $ denotes the Floquet quasienergy and $ \ket{\Phi(\tau)} = \ket{\Phi(\tau + T)} $ is the Floquet eigenvector. By applying the Fourier decomposition to both the Hamiltonian $ H(\tau) = \sum_n \me^{-\mi n \omega \tau} H_n $ and  $ \ket{\Phi(\tau)} = \sum_n \me^{-\mi n \omega\tau}\ket{\Phi^n} $, the time-dependent Schrödinger equation is mapped to an infinite-dimensional eigenvalue equation in the extended Hilbert space:
\begin{align}\label{Eq.3}
    \begin{split}
        \sum_{m} (H_{n-m} - m\omega\delta_{mn}) \ket{\Phi_{\alpha}^m} = \epsilon_{\alpha} \ket{\Phi_{\alpha}^n}.
    \end{split}\tag{3}
\end{align}
Throughout this work, we focus on the high-frequency regime where direct electron transitions mediated by photon absorption or emission are strongly suppressed. In this off-resonant limit, the time-dependent system can be effectively described by a static Floquet Hamiltonian, derived as~\cite{Bukov04032015,Eckardt2015}:
\begin{align}\label{Eq.4}
    \begin{split}
        {\cal H}_{\text{eff}} = {\cal H}_{0} + \sum_{n\neq0} \frac{ \left[ {\cal H}_{-n}, {\cal H}_{n} \right] }{ 2 n \omega } + {\cal O}(\omega^{-2}).
    \end{split}\tag{4}
\end{align}
While the truncation order in numerical calculations is typically determined by convergence criteria, we focus here on the leading-order contributions, retaining terms up to the cutoff $ n = 1 $.


\textit{Floquet Engineering of Chiral Topological Phases in AM/s-wave SC}.--- 
To explore the system's response under nonequilibrium conditions, we explicitly consider $s$-wave pairing by fixing $\Delta_0 = 0.3$ and $\Delta_1 = 0$ for the subsequent analysis.
In the absence of an external light field, tuning the AM strength $J_\text{AM}$ places the system in either a topologically trivial phase or a weak TSC phase~\cite{PhysRevLett.133.106601}, and we adopt these as our initial states.
Detailed discussions on the intrinsic band structure and momentum-dependent spin splitting in the absence of the EPL are provided in the Supplemental Material \cite{SM}.
Figure~\ref{FIG2}(a) and (b) display the topological phase diagrams in the $\theta$-$A$ parameter space for $J_\text{AM} = 0.245$ and $J_\text{AM} = 0.26$, respectively. Focusing on the $J_\text{AM} = 0.245$ case, the introduction of EPL drives the system from a trivial phase into a strong TSC phase ($\mathcal{N} = \pm 1$). Moreover, fixing $A$ and varying $\theta$ reveals multiple topological phase transitions where the Chern number evolves $1 \to 0 \to -1$. 
Each transition is accompanied by a band inversion. To elucidate the physical mechanism governing this evolution, we analyze the topological phase transitions along the cut defined by $A=0.8$ [indicated by the red dashed line in Fig.~\ref{FIG2}(a)]. This series of phase transitions is due to the anisotropic response of the X and Y valleys to the EPL field.
As shown in Fig.~\ref{FIG3}(a), for $\theta = 0.40\pi$, the system exhibits a Chern number of $\mathcal{N}=1$. As $\theta$ decreases to $0.31\pi$, the band gap at the X valley closes (projected onto the $\bar{\Gamma}$ point) [Fig.~\ref{FIG3}(b)]. Upon further decreasing $\theta$, the X valley gap reopens, driving the system into a trivial state with $\mathcal{N}=0$ [Fig.~\ref{FIG3}(c)]. Subsequently, at $\theta = 0.18\pi$, the gap at the Y valley closes [Fig.~\ref{FIG3}(d)]. Finally, as $\theta$ continues to decrease, a band inversion occurs at the Y valley, resulting in $\mathcal{N}=-1$ [Fig.~\ref{FIG3}(e)].
Turning to the case of $J_\text{AM}=0.26$, where the system originates from the weak TSC phase, the introduction of EPL similarly drives a transition toward a strong TSC phase. We focus on the evolution along the cut at $\theta = 0.225\pi$ [red dashed line in Fig.~\ref{FIG2}(b)]. By varying the amplitude $A$, we observe distinct edge state behaviors. At $A=0.1$, the system hosts two Majorana edge modes of opposite chirality. Notably, the Chern number remains $\mathcal{N}=0$, as shown in Fig.~\ref{FIG3}(f). As $A$ increases to $0.46$, the band gap at the Y valley closes [Fig.~\ref{FIG3}(g)]. Upon further increasing $A$, the gap at the Y valley reopens, yielding a phase with $\mathcal{N}=1$ [Fig.~\ref{FIG3}(h)]. Finally, with a continued increase in $A$, the X valley undergoes a gap closing and reopening process, returning the system to a state with $\mathcal{N}=0$, as illustrated in Figs.~\ref{FIG3}(i) and (j).

In summary, we have demonstrated that the application of EPL consistently drives the system into a strong TSC phase characterized by an odd Chern number, regardless of whether the initial static phase is topologically trivial or a weak TSC state. Furthermore, our results highlight that the system allows for versatile topological control, where multiple phase transitions can be precisely realized by tuning the polarization angle $\theta$ and the light amplitude $A$. To further illustrate the versatility of the system, a comprehensive discussion on the topological phase transitions cooperatively driven by the altermagnetic exchange strength and the light field amplitude is presented in the SM~\cite{SM}

\begin{figure}{t}
    \centering
    \includegraphics[width=0.48\textwidth]{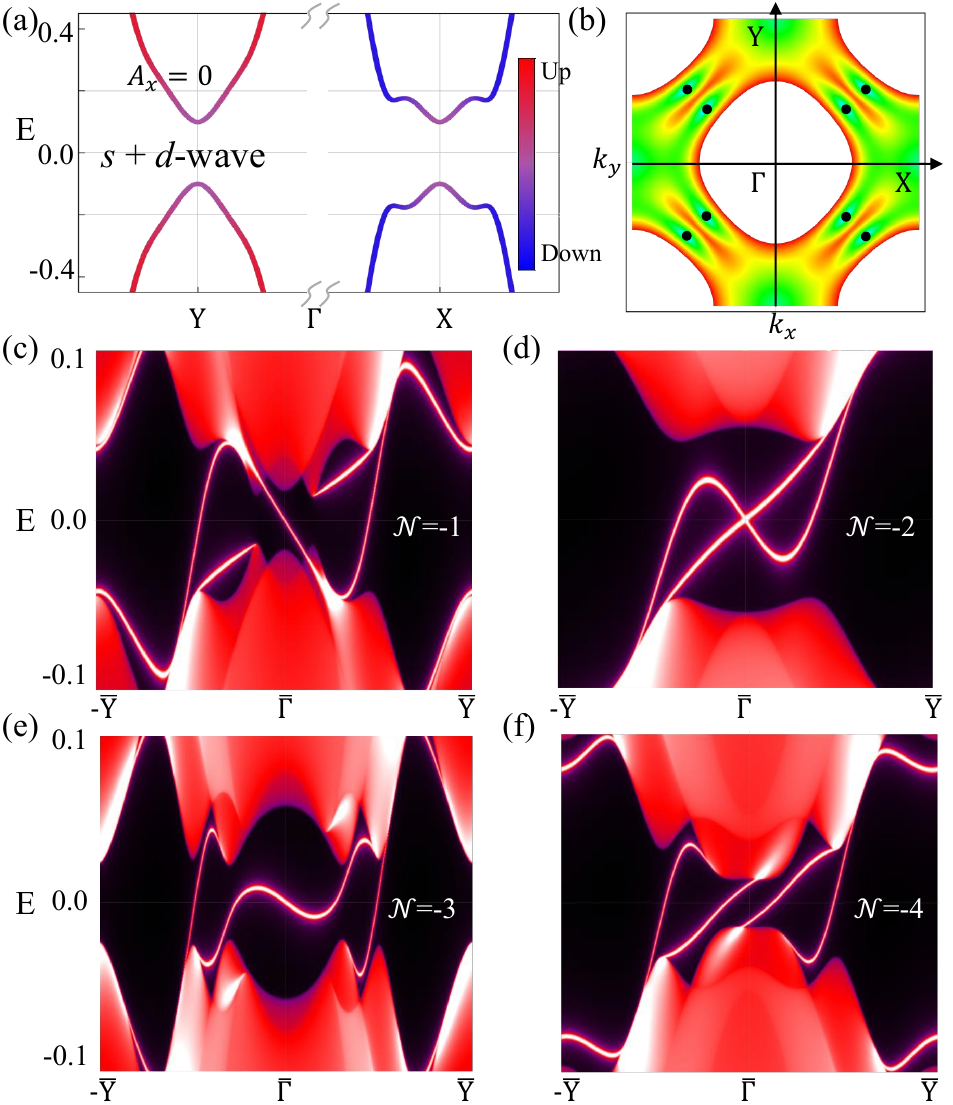}
    \caption{Topological phases in an AM/$ s+d $-wave SC system.
    (a) Spin-resolved band structures along the high-symmetry path $ \text{Y} $-$ \Gamma $-$ \text{X} $ in the static limit ($ A_x = 0, J_\text{AM} = 0.3 $).
    (b) Illustration of the distribution of bulk nodes in the Brillouin zone. 
    (c)-(f) Topological edge state spectra of the AM/$ s+d $-wave SC heterostructure along the surface Brillouin zone path $ \bar{\text{Y}} $-$ \bar{\Gamma} $-$ \bar{\text{Y}} $, driven by an EPL with a varying light amplitude $ A_x $ and exchange strength $ J_\text{AM} $. The specific parameters and resulting phases are: 
    (c) $ \mathcal{N}  = -1 $ with $ J_\text{AM} = 0.6 $ and $ A_x = 0.98 $; 
    (d) $ \mathcal{N}  =- 2 $ with $ J_\text{AM} = 0.4 $ and $ A_x = 1.05 $; 
    (e) High-Chern-number state $ \mathcal{N} = -3 $ with $ J_\text{AM} = 0.4 $ and $ A_x = 0.7 $; 
    and (f) $ \mathcal{N} =- 4 $ with $ J_\text{AM} = 0.4 $ and $ A_x = 0.90 $. 
    Other model parameters are fixed at $ \mu = 0, \lambda_\text{R} = 0.3 $, and $ t = 1 $ for all plots.
    \label{FIG4}}
\end{figure}

\textit{Floquet Engineering of Chiral Topological Phases in AM/s+d-wave SC}.--- 
Next, we consider the $ s+d $-wave pairing by setting $ \Delta_0 = 0.1 $ and $ \Delta_1 = 0.3 $. For computational convenience, we set $A_y/A_x = 2$ in the following analysis. Under these conditions, we observe the emergence of gapless bulk nodes in the system. The resulting band structure and the distribution of these nodes in the momentum space are presented in Figs.~\ref{FIG4}(a) and (b), respectively. Upon introducing an EPL, a full energy gap opens at the bulk nodes, extending across the entire Brillouin zone. 

Furthermore, we found that the interplay between AM and the optical drive uncovers remarkably rich topological phases. 
We observe that the system undergoes a cascade of sequential phase transitions, evolving into high-Chern-number states ranging from $ \mathcal{N} = -1 $ up to $ \mathcal{N} = -4 $, as visualized in Figs.~\ref{FIG4}(c)-(f). Crucially, these topological invariants strictly obey the bulk-boundary correspondence: the phases with $ \mathcal{N} = -3 $ and $ \mathcal{N} = -4 $ signify the emergence of three and four chiral Majorana edge modes within the bulk gap, respectively. We thus conclude that the FCTSC phases in our driven AM system can be effectively engineered to host a highly tunable number of propagating gapless edge channels.

\textit{Conclusion}.--- 
In summary, we have theoretically investigated the topological phase transitions in an AM/SC heterostructure driven by EPL. To provide a comprehensive understanding, we conducted a comparative study considering two distinct superconducting pairings: $ s $-wave and $ s+d $-wave.
By systematically analyzing the interplay between the AM strength and the external periodic driving, we demonstrated that the system can be driven into multiple FCTSC phases. 
For example, under the AM/$s$-wave superconducting pairing, the system can evolve from either a topologically trivial phase or a weak TSC phase into a strong TSC state characterized by an odd Chern number $\mathcal{N}$.
Furthermore, within the vanishing BdG Chern number regime ($ \mathcal{N}= 0 $), our analysis reveals a subtle distinction between a trivial state and a weak FCTSC. The latter is notably characterized by the presence of two Majorana edge modes of opposite chirality, protected by translational symmetry. In comparison, for the $ s+d $-wave pairing, the topological classification spans a much wider range, accessing high BdG Chern numbers from $ \mathcal{N} = -1 $ to $ \mathcal{N} = -4 $ by adjusting the light amplitude and AM strength. Overall, the extensive tunability of these topological phases underscores the potential of Floquet engineering for manipulating exotic quantum states within AM/SC heterostructures.

\emph{Acknowledgments.}---
This work was supported by the National Key Research and Development Program of the Ministry of Science and Technology of China (Grant No. 2025YFA1411303), the National Natural Science Foundation of China (NSFC, Grants No. 12595330, No. 92365101, No. 12474151, No.  12547101, No. 12447141), and the Natural Science Foundation of Chongqing (Grant No. 2023NSCQ-JQX0024 and CSTB2023NSCQ-MSX0476).

\bibliography{ref2}

\end{document}